\documentclass[12pt]{article}
\usepackage{graphicx}
\usepackage{color}
\usepackage{cite}

\def\Xiccpp{\hbox{$\Xi_{cc}^{++}$}}
\def\Xiccp{\hbox{$\Xi_{cc}^{+}$}}
\def\stat{\hbox{\ (stat.)}}
\def\syst{\hbox{\ (syst.)}}
\def\mevcc{\hbox{\ MeV}}

\def\beq{\begin{equation}}
\def\eeq{\end{equation}}
\def\eqref#1{(\ref{#1})}
\def\bea{\begin{eqnarray}}
\def\eea{\end{eqnarray}}

\def\URLtilde{\lower0.2em\hbox{$\tilde{\phantom{a}}$}}
\def\mycomm#1{\hfill\break\strut\kern-3em{\color{red}\tt ====> #1
\color{black}}\hfill\break}

\newcount\timecount
\newcount\hours \newcount\minutes  \newcount\temp \newcount\pmhours
\hours = \time
\divide\hours by 60
\temp = \hours
\multiply\temp by 60
\minutes = \time
\advance\minutes by -\temp
\def\hour{\the\hours}
\def\minute{\ifnum\minutes<10 0\the\minutes
\else\the\minutes\fi}
\def\clock{
\ifnum\hours=0 12:\minute\ AM
\else\ifnum\hours<12 \hour:\minute\ AM
\else\ifnum\hours=12 12:\minute\ PM
\else\ifnum\hours>12
\pmhours=\hours
\advance\pmhours by -12
\the\pmhours:\minute\ PM
\fi
\fi
\fi
\fi
}

\def\monthname{\relax\ifcase\month 0/\or January\or February\or
March\or April\or May\or June\or July\or August\or September\or
October\or November\or December\else\number\month/\fi}

\def\bold#1{\setbox0=\hbox{$#1$}     \kern-.025em\copy0\kern-\wd0
\kern.05em\copy0\kern-\wd0
\kern-.025em\raise.0433em\box0 }

\begin{document}
\setcounter{footnote}{1}
\vskip1.0cm

\centerline{\large \bf 
LHCb gets closer to discovering the second doubly charmed baryon%
\footnote{To appear in {\em News \& Views}, Sci. China-Phys. Mech. Astron. 63(2): 221064 (2020).}
}
\bigskip

\centerline{Marek Karliner$^a$\footnote{{\tt marek@tauex.tau.ac.il}}
 and Jonathan L. Rosner$^b$\footnote{{\tt rosner@hep.uchicago.edu}}}
\medskip
\medskip

\centerline{$^a$ {\it School of Physics and Astronomy}}
\centerline{\it Tel Aviv University, Tel Aviv 69978, Israel}
\medskip

\centerline{$^b$ {\it Enrico Fermi Institute and Department of Physics}}
\centerline{\it University of Chicago, 5640 S. Ellis Avenue, Chicago, IL
60637, USA}
\bigskip
\strut

Recently the LHCb Collaboration published the results of a
search for the doubly charmed baryon \,\Xiccp\ \cite{Aaij:2019jfq}.
No significant signal is seen in the mass range from 3.4 to 3.8 GeV.  To put
this result in context, the \Xiccpp\ baryon was seen by LHCb in decay modes
$\Lambda_c K^- \pi^+ \pi^+$ \cite{LHCb-PAPER-2017-018} (2017) and $\Xi_c^+
\pi^+$ \cite{LHCb-PAPER-2018-026} (2018).  The weighted average of the \Xiccpp
\,\ mass is $3621.24\pm0.65\stat \pm0.31\syst\mevcc$\cite{LHCb-PAPER-2018-026}.

The \Xiccpp\ and \Xiccp\ have the quark content $ccu$ and $ccd$, respectively.
Under the isospin symmetry of the strong interactions they form an isodoublet,
like the proton and the neutron. Isospin breaking in hadron masses is a very
small effect \cite{Karliner:2019lau}.  Consequently we have firm reasons to
expect that the $\Xiccpp-\Xiccp$\, mass difference is quite small, ${\cal O}
(1.5)$ MeV \cite{Karliner:2017gml}. The production rates of \,\Xiccpp\,\ and
\,\Xiccp\,\ should be similar, as the bottleneck --- the production of the $cc$
diquark --- is the same in both cases.  Consequently we {\em know} \,\Xiccp\,\
exists in the vicinity of 3620 MeV.  A claimed $\Xiccp$ at 3518.7$\pm$1.7
MeV \cite{Mattson:2002vu,Ocherashvili:2004hi} is unlikely to be the isospin
partner of the established $\Xiccpp$, and has not been confirmed by any other
experiment.

The search was a ``blind analysis", i.e., it was performed with the whole
procedure defined before inspecting the data in the 3400 to 3800 MeV mass
range.  A search for a \Xiccp\ signal was performed and the significance of
the signal as a function of the \Xiccp\ mass was evaluated.  If the global
significance,
after considering the look-elsewhere effect, was found to be above 3$\sigma$,
the \Xiccp\ mass was measured; otherwise, upper limits were set on the
production rates for different CM energies.

As can be seen from Fig.~2 in Ref.~\cite{Aaij:2019jfq}, the data exhibit
several peaks, but the most significant one occurs just where it is expected.
The largest local significance, corresponding to 3.1$\sigma$ (2.7$\sigma$ after
considering systematic uncertainties), occurs around 3620 MeV. However, the
look-elsewhere effect \cite{Gross:2010qma}, intrinsic to the LHCb search
procedure, reduces this to 1.7$\sigma$. We believe that in this
case the look-elsewhere effect may be overstated, because the peak shows up
nearly (but not precisely) where expected.  The result of a fit, as given in
the Supplementary Material of Ref.\ \cite{Aaij:2019jfq}, is $M(\Xiccp) =
3623.4\pm 1.7$ MeV, a bit {\it larger} than $M(\Xiccpp)$, in contrast to the
prediction of Ref.\ \cite{Karliner:2017gml} and nearly all the others quoted
there which find $M(\Xiccp)$ less than but within a few MeV of $M(\Xiccpp)$.

The upper limit on $\Xiccp$ production (or the significance of a signal)
increases with shorter assumed lifetime, as seen in Table 6 and Fig.\ 6 of Ref.\
\cite{Aaij:2019jfq}.  As a result of the internal $c d \to s u$ process in the
decay of $\Xiccp$, its lifetime is several times shorter than that of
$\Xiccpp$:  For example, Ref.\ \cite{Karliner:2014gca} finds $\tau(\Xiccp) =
53$ fs, and $\tau(\Xiccpp) = 185$ fs.  (The latter was measured by LHCb to be
$256^{+24}_{-22} \pm 14$ fs \cite{Aaij:2018wzf}.)

The validity of the prediction of $M(\Xiccpp)$ \cite{Karliner:2014gca} and the
signal of its isospin partner not far from {\it its} predicted mass
\cite{Karliner:2017gml} lend credence to an estimate of the mass of the
$c c \bar u \bar d$ tetraquark using similar methods, which finds this state
to have a mass of $3882 \pm 12$ MeV \cite{Karliner:2017qjm} and hence unstable
with respect to strong decay.  The cross section for $c c \bar u \bar d$
tetraquark production is expected \cite{Karliner:2017qjm} to be somewhat, but
not much, smaller than the cross section for production of \,\Xiccpp\,\ and
\,\Xiccp\, baryons.  Thus the new LHCb results provide additional motivation
for continuing the search for the $c c \bar u \bar d$ tetraquark.

In summary, the data contain a 2.7$\sigma$ hint of the \Xiccp\,\ signal at 
a mass consistent with predictions based on the measured \Xiccpp\,\ mass 
and isospin symmetry.  More data are needed to exclude the possibility that this
is a statistical fluctuation.

\end{document}